\documentclass[letter]{aa} % for the letters 
\usepackage{graphicx,comment}
\usepackage{txfonts}
\begin{document} 

   \title{Turbulent fragmentation as the primary driver of core formation in Polaris Flare and Lupus I
   \thanks{{\it Herschel} is an ESA space observatory with science instruments provided by European-led Principal Investigator consortia and with important participation from NASA.}
   }

   \subtitle{}

\titlerunning{Turbulent fragmentation in Polaris Flare and Lupus I}
\authorrunning{Ishihara et al.}

   \author{Kousuke Ishihara\inst{1,2},
   Fumitaka Nakamura\inst{1,2,3},
          Patricio Sanhueza\inst{1,4},
          \and
          Masao Saito\inst{1,2}
}
   \institute{
   National Astronomical Observatory of Japan, National Institutes of Natural Sciences, 2-21-1 Osawa, Mitaka, Tokyo 181-8588, Japan
   \label{NAOJ}
   \and
    Department of Astronomical Science, SOKENDAI (The Graduate University for Advanced Studies), 2-21-1 Osawa, Mitaka, Tokyo 181-8588, Japan
    \label{SOKENDAI}
    \and
    Department of Astronomy, Graduate School of Science, The University of Tokyo, 7-3-1 Hongo, Bunkyo-ku, Tokyo 113-0033, Japan
    \label{TODAI}
    \and
    Department of Earth and Planetary Sciences, Institute of Science Tokyo, Meguro, Tokyo, 152-8551, Japan
    \label{TOKODAI}}

% \abstract{}{}{}{}{} 
% 5 {} token are mandatory
 
  \abstract
   {Stars form from dense cores in turbulent molecular clouds. According to the standard scenario of star formation, dense cores are created by cloud fragmentation. However, the physical mechanisms driving this process are still not fully understood from an observational standpoint.}
   {Our goal is to investigate the process of cloud fragmentation using observational data from nearby clouds. Specifically, we aim to examine the role of self-gravity and turbulence, both of which are key to the dynamical evolution of clouds.
  }
   {We applied \texttt{astrodendro} to the {\it Herschel} H$_2$ column density maps to identify dense cores and determine their mass and separation in two nearby low-mass clouds: the Polaris Flare and Lupus I clouds. We then compared the observed core masses and separations with predictions from models of gravitational and turbulent fragmentation. In the gravitational fragmentation model, the characteristic length and mass are determined by the Jeans length and Jeans mass. For turbulent fragmentation, the key scales are the cloud’s sonic scale and its corresponding mass.
   }
   {The average core masses are estimated to be 0.242 $M_\odot$ for Lupus I and 0.276 $M_\odot$ for the Polaris Flare. The core separations peak at about $2-4 \times 10^4$ au ($\approx$ 0.1 -- 0.2 pc) in both clouds. These separations are significantly smaller than the Jeans length but agree well with the cloud sonic scale.
    Additionally, the density probability distribution functions of the dense cores follow log-normal distributions, which is consistent with the predictions of turbulent fragmentation.}
   {These findings suggest that the primary process driving core formation in the observed low-mass star-forming regions is not gravitational fragmentation but rather turbulent fragmentation. We found no evidence that filament fragmentation plays a significant role in the formation of dense cores.
}


   \maketitle
%
%-------------------------------------------------------------------

\section{Introduction}

Molecular clouds fragment into dense, compact cores, with gravitationally bound cores being widely recognized as the primary sites of protostellar formation. The process of gravitational fragmentation is considered key to core creation in molecular clouds \citep[e.g.][]{Larson_1985,Andre_2010}. Recent observations from the Atacama Large Millimeter/submillimeter Array (ALMA) appear to have provided empirical support for this concept. These observations show that the core separations within high-mass, cluster-forming molecular clumps are in reasonable agreement with the predictions of thermal Jeans fragmentation \citep{klaassen18,Beuther_2022,Lu_2020,Zhang_C_2022_ATOMS,Ishihara_2024,sanhueza19,morii24}.
This suggests that thermal gravitational, or thermal Jeans, fragmentation governs core formation in high-mass regions.

In particular, \citet{Ishihara_2024} used 1.3mm dust continuum images of 30 high-mass cluster-forming clumps observed as part of the DIHCA 
%survey
(Digging into the Interior of Hot Cores with ALMA) survey,
%\textcolor{red}{\citep[Digging into the Interior of Hot Cores with ALMA, see also][]{Olguin_2021,Olguin_2022,Taniguchi_2023,Olguin_2023}},
identifying around $10^3$ dense cores. Their results show that the typical core separation is about 7800 au, which is consistent with the thermal Jeans length of the parent clumps:
$ L_J= c_s \sqrt{\pi/G\rho} $,
where $c_s$ is the sound speed, $G$ is the gravitational constant, and $\rho$ is the density.
The corresponding core mass, given by the Jeans mass, is $M_J= 4\pi \rho (L_J/2)^3/3 $.

\begin{table*}[ht]
\centering
%\scriptsize
\caption{Physical properties of target clouds. \label{tab:sources}}
\begin{tabular}{lccccccccccc}
\hline
Cloud & Distance & $M_{\rm{cl}}$ & $R_{\rm{cl}}$ & $\langle T_{\rm{cl}}\rangle$ & $\langle N_{\rm{cl}}\rangle$ & $\langle n_{\rm{cl}}\rangle $ & $M_{\rm J, cl}$ & $L_{\rm J, cl}$ & $N_{\rm J, cl}$ & $N_{\rm proto}$  \\
 & (pc) & ($M_\odot$) & (pc) & (K) & ($\times 10^{20} \rm{cm}^{-2}$) & ($\times 10^3 \rm{cm}^{-3}$) & ($M_\odot$) & (pc) & \\
\hline
Polaris & 355\tablefootmark{a} & 1260 & 4.6 & 15.1 & 8.4 & 0.04 & 77.1 & 3.8 & 16.4 & 0 \\
Lupus I & 182\tablefootmark{b} & 1080 & 3.6 & 16.6 & 11.9 & 0.08 & 65.7 & 3.0 & 16.4 & 9$^b$ \\
\hline
\end{tabular}
\tablefoot{$^{(a)}$\citet{panopoulou22} for Polaris, and $^{(b)}$\citet{Benedettini_2018} for Lupus I.
See also the Appendix \ref{sec:distance}.}
\label{tab:targets}
\end{table*}

These results suggest that thermal Jeans fragmentation plays a significant role in core formation within dense, cluster-forming regions. However, it remains unclear whether the same dynamics apply in other environments. In several nearby molecular clouds, such as Orion A and $\rho$-Ophiuchus, most dense cores have been found to be gravitationally unbound and  pressure-confined \citep{maruta10,Kirk_2017,takemura23}. The mechanisms responsible for the formation of these unbound cores remain uncertain.

This paper focuses on nearby low-mass clouds, specifically the Polaris Flare (hereafter Polaris) and Lupus I, where the role of cloud self-gravity is relatively diminished compared to the influence of supersonic turbulence, which may even dominate over self-gravity. To achieve a deeper understanding of core formation in various cloud environments, we apply methods similar to those used by \citet{Ishihara_2024} to investigate the physical processes governing core formation in these less dense, low-mass molecular clouds.

We consider three key fragmentation mechanisms: thermal Jeans fragmentation, turbulent Jeans fragmentation, and turbulent fragmentation. In thermal Jeans fragmentation, the balance between thermal pressure and self-gravity determines how gravitationally bound structures form. The characteristic mass and length scales in this scenario are the Jeans mass ($M_J$) and Jeans length ($L_J$), which depend on the cloud's temperature and density \citep{Jeans_1928,Larson_1985}.

Turbulent Jeans fragmentation is a variation of thermal Jeans fragmentation, whereby turbulent pressure supports the cloud, leading to larger characteristic mass and length scales than in the thermal case. However, this model does not account for the full properties of cloud turbulence and is only applicable to larger structures for which turbulent motions can be approximated as isotropic \citep[e.g.][]{maclow04,palau15}.

In turbulent fragmentation, dense cores form through the dynamical compression of local turbulence \citep{padoan02}. The density probability distribution function (PDF) in turbulent clouds tends to follow a log-normal distribution due to repeated compression \citep{Padoan_2020}. In a turbulent medium with a power-law energy spectrum, there are no inherent characteristic mass and length scales. However, in the interstellar medium, the power law breaks at the sonic scale, 
$l_S$, where turbulent velocity dispersions become comparable to the sound speed \citep{Federrath_2012}. For nearby molecular clouds, this scale is typically measured to be 
$l_S\sim $ 0.1 pc ($\approx 20000$ au).  Therefore, the sonic scale, 
$l_S$, may represent the characteristic length for turbulent fragmentation, with the corresponding mass ($M_S$) estimated as
$M_S\approx 4 \pi \left<\rho\right> (l_S/2)^3/3$, respectively, 
where $\left<\rho\right>$ is the average cloud density. Since these three fragmentation mechanisms predict different characteristic mass and length scales, we aim to constrain the dominant fragmentation process in these nearby clouds by deriving core masses and separations from observational data.

The paper is structured as follows: 
Sect. \ref{sec:obs} details the data used in this study. In Sect. \ref{sec:result1}, we present the results of core identification and their physical properties. We then compare these core properties with predictions from gravitational fragmentation and turbulent fragmentation models in Sect. \ref{sec:grav} and \ref{sec:turb}, respectively, and discuss the physical processes involved in core formation in these molecular clouds in Sect. \ref{sec:conclusion}.
\vspace{-2mm}

%--------------------------------------------------------------------
\section{Targets, data, methods, and core identification}
\label{sec:obs} 
%-------------------------------------- Two column 

\begin{table*}[ht]
\centering
%\scriptsize
\caption{Physical properties of identified cores.}
\begin{tabular}{lccccccccc}
\hline
Cloud & $\mathcal{N}_{\rm core}$ & $\left< M_{\rm core} \right>$ & $\left< R_{\rm core} \right>$ & $\left< T_{\rm core} \right>$ & $\left< N_{\rm core} \right>$ & $\left< n_{\rm core} \right>$ & \multicolumn{2}{c}{$\alpha_{\rm BE}$} & $\left< S_{\rm 3D} \right>$ \\
 &  & ($M_\odot$) & (pc) & (K) & ($10^{21}$ cm$^{-2}$) & ($10^4$ cm$^{-3}$) & min & max & (pc) \\
\hline
Polaris & 407 & 0.276 & 0.077 & 14.7 & 2.02 & 18.71 & 0.7 & 20.0 & 0.27 \\
Lupus I & 432 & 0.242 & 0.037 & 15.1 & 6.42 & 27.83 & 0.2 & 35.2 & 0.13 \\
\hline
\end{tabular}
\label{tab:cores}
\end{table*}

\begin{figure*}[htb!]
%\begin{figure*}[H]
\centering
\includegraphics[width= 0.85\linewidth]{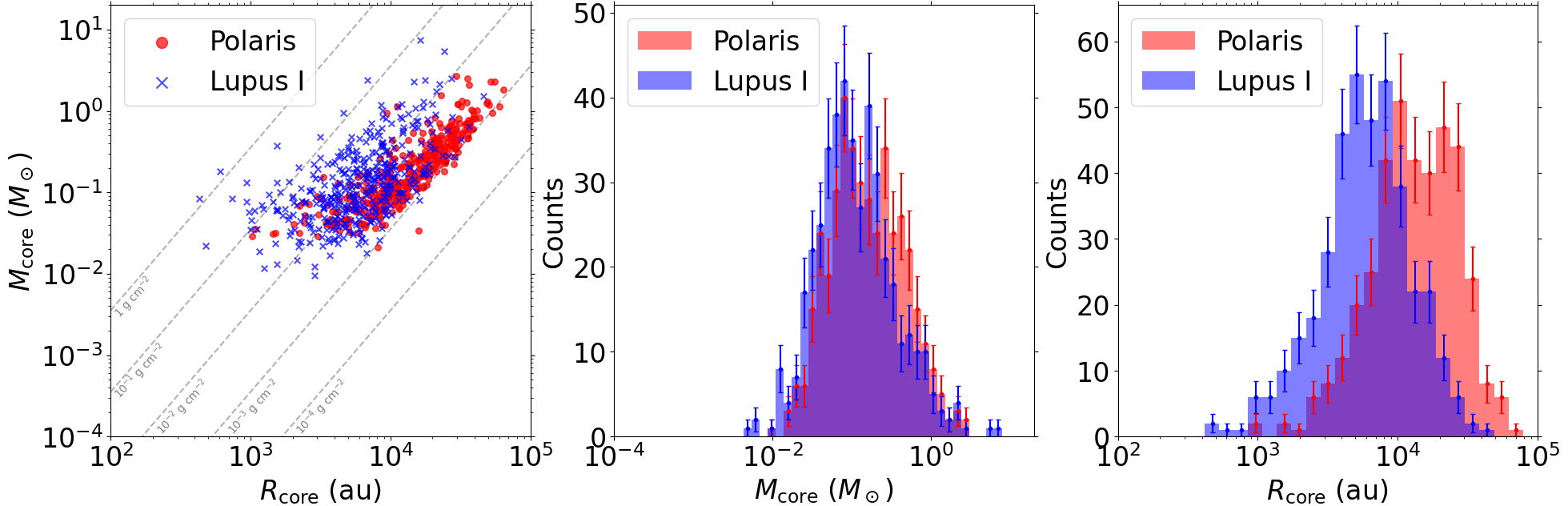}
\caption{Core mass versus radius diagram ({\it left}) and histograms of core mass (${\it middle}$) and core radius (${\it right}$) for Polaris and Lupus I by \texttt{astrodendro}. The red circle and blue cross symbols represent the cores in Polaris and Lupus I, respectively.
The error bar was calculated by Poisson statistics.
\label{fig:M-R_hsit}}
\end{figure*}

From the {\it Herschel} Gould Belt Survey \citep{Andre_2010}, we have chosen two clouds detailed in Tab. \ref{tab:targets}: Polaris and Lupus I. High-resolution (18\arcsec) column density maps of these clouds were obtained from the {\it Herschel} Gould Belt Survey's website\footnote{\url{http://www.herschel.fr/cea/star-formation/en/Phocea/Vie_des_labos/Ast/ast_visu.php?id_ast=63}}. At the distances to the clouds (355 pc for Polaris, and 182 pc for Lupus I), the 18\arcsec \, resolution corresponds to 6390 au and 2700 au, respectively (see Fig. \ref{fig:herschel maps}).
The average density was computed by dividing the total cloud mass by the volume of a sphere with a radius of $R_{\rm cl} = \sqrt{{\rm Area}/\pi}$, where the cloud area was determined by that of the trunk identified by \texttt{astrodendro}.
As is shown below, these clouds exhibit a sufficient number (several hundred or more) of identified dense cores, using \texttt{astrodendro}.

Polaris and Lupus I are categorized as a quiescent region and a low-mass star-forming region, respectively, according to \citet{Schneider_2022}. 
In Lupus I, nine protostars are identified \citep{Benedettini_2018}, whereas Polaris has not formed protostars yet. 
Both clouds share similar average column densities and temperatures, with $\left< N_{\rm cl} \right> \approx 8-12\times 10^{20} , \rm cm^{-2}$ and $\left<T_{\rm cl} \right>\approx 15 , \rm K$. As a result, the Jeans mass of the clouds is estimated to be around 70--80 $M_\odot$, with a corresponding Jeans length of about 3--4 pc.
However, the degree of density concentration is different (see Fig. \ref{fig:herschel maps}).
The molecular gas in Polaris appears to be more uniformly and spatially extended.
The gravitational boundness of the cloud is likely weaker, given that the virial parameters for the Polaris and Lupus I clouds are estimated to be about 3.1 and 2.1, respectively, based on an observed velocity dispersion of 1.51 km s$^{-1}$ and 1.46 km s$^{-1}$ \citep{Spilker_2022}.

We applied \texttt{astrodendro} \citep{Rosolowsky_2008} to the column density maps with the threshold value of \texttt{min\_value} ($S_{\mathrm{min}} = 5\sigma $), the minimum step of \texttt{min\_delta} ($\delta_{\mathrm{min}}=3 \sigma $), and the minimum pixel number of 
\texttt{min\_npix} ($\theta_{\mathrm{min}}$ was set to the number of pixels equivalent to the beam area).
Here, we adopted 1-$\sigma$ noise levels of $7.67 \times 10^{19}$ cm$^{-2}$ for Polaris and $8.02 \times 10^{19}$ cm$^{-2}$ for Lupus I, respectively (for this choice of parameters, see Appendix \ref{sec:dependence}).
The \texttt{astrodendro} identifies three structures: a leaf, branch, and trunk. 
We defined the minimum structure of the `leaf' as a dense core and derived the core mass by subtracting the background gas components, which are defined as the column density of the corresponding trunk component.
Following \citet{takemura21b}, 
we used additional condition to select cores: the peak column density of the core should be more than 2$\times$ min\_value \citep[see also][]{shimajiri15}.
The positions of the identified cores are indicated in Fig. \ref{fig:herschel maps}.
Our result show that in Lupus I, nine protostars are identified 
\citep[compared with the catalogue by][]{Benedettini_2018},
%(compared with the catalogue by \citet{Benedettini_2018}), 
whereas Polaris has not formed protostars yet.  This is consistent with previous studies \citep{Benedettini_2018, ward-thompson10}. A more detailed comparison is given in Appendix \ref{sec:comparison}. The core identification also depends on the parameters chosen. In Appendix \ref{sec:dependence}, we examine the dependence of \texttt{astrodendro} parameters. In Appendix \ref{sec:comparison}, we  also compare the core properties with the ones of \citet{Benedettini_2018}, who used \texttt{getsources}.

Then, we applied the nearest neighbour search (NNS) method 
to derive the core separations on the 2D H$_2$ column density image
\citep{Knuth_1973}.
The core separations ($S_{3D}$) were obtained by multiplying a correction factor of $4/\pi$ (see Appendix \ref{sec:nns}). 
%\vspace{-2mm}

%\section{Results}
\section{Physical properties of dense cores}
\label{sec:result1}
We identified the dense cores; their physical properties are summarized in Tab. \ref{tab:cores}.  
Here, the core radius, $R_{\rm core}$, was calculated as half of the geometric mean of the FWHM$_{\rm major}$ and FWHM$_{\rm minor}$ provided by \texttt{astrodendro}; specifically, $R_{\rm core} = 0.5 \sqrt{{\rm FWHM}_{\rm major} \times {\rm FWHM}_{\rm minor}}$.
The core radius was then beam-deconvolved.
The core density, $n_{\rm core}$, was estimated under the assumption of spherical symmetry using $n_{\rm core} = M_{\rm core}/ (4\pi R_{\rm core}^3/3)$.
The identified cores were classified into two groups: gravitationally bound cores and unbound cores, based on the Bonnor-Ebert ratio ($\alpha_{\rm BE} =  M_{\rm BE}^{\rm crit}/M_{\rm core}$), which compares the core mass ($M_{\rm core}$) to the critical Bonnor-Ebert mass (see Sect. 4.1 of \citet{Konyves_2010} for the definition and discussion of $M_{\rm BE}^{\rm crit}$.). The number of identified cores, $\mathcal{N}_{\rm core}$, in each cloud significantly exceeds the cloud Jeans number ($N_{J,cl}=M_{\rm cl}/M_{J, cl}$), suggesting that cloud structures are hierarchical and that cores primarily form in denser regions. This implies that core formation may not be predominantly controlled by gravitational fragmentation, or may be a combination of both hierarchical structures and other processes.

In Polaris, the cores are somewhat larger than the ones in Lupus I (see also Fig. \ref{fig:M-R_hsit}). The average core radii are estimated to be 0.077 pc for Polaris and 0.037 pc for Lupus I, while the average core masses are 0.28 M$_\odot$ and 0.24 M$_\odot$, respectively. The core density tends to be higher in Lupus I (see also the left panel of Fig. \ref{fig:M-R_hsit}).

To judge the gravitational boundness of cores, we derived $\alpha_{\rm BE}$ \citep{Andre_2014}.
In Polaris, only seven gravitationally bound cores were found, whereas about 18 \% of the cores ($\approx 80$) in Lupus I are gravitationally bound. This is consistent with the fact that Lupus I exhibits star formation activity in its denser regions.

\section{Comparison with gravitational fragmentation}
\label{sec:grav}

\begin{figure*}[htb!]
%\begin{figure*}[H]
\centering
\includegraphics[width=0.95\linewidth]{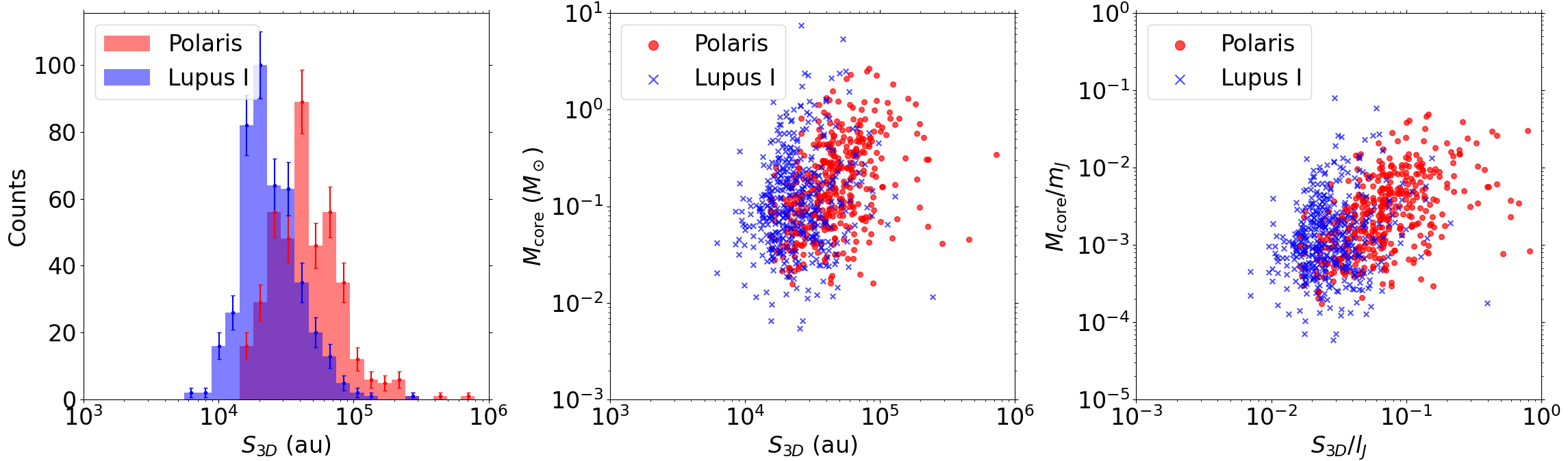}
\caption{Histogram of core separation (${\it left}$) and core mass versus separation diagrams in physical scale ({\it middle}) and normalized scale (${\it right}$).
The red circle and blue cross symbols represent the cores in Polaris and Lupus I, respectively. For the normalized scale, the core mass was divided by the local thermal Jeans mass ($m_J$) and the core separation was divided by the local thermal Jeans length ($l_J$), where $m_J$ and $l_J$ are estimated in the trunks that include the corresponding cores (leaves).
The error bar was calculated by Poisson statistics.
\label{fig:M-separation_hsit}}
\end{figure*}

The upper panel of Fig. \ref{fig:M-separation_hsit} shows the histogram of core separations for the two clouds.
Interestingly, the distribution of core separation is quite similar to each other, peaking at $\sim 1-2 \times 10^4 $ au ($\approx 0.1-0.2 $ pc).
Given that the Jeans length for these clouds is about 4 pc (= 8 $\times 10^5$ au), 
the peak separation is about 20--40 times smaller than the cloud Jeans length.
The density at which the Jeans length coincides with the peak separation is estimated to be of the order of $10^5$ cm$^{-3}$.
This density is significantly higher than what is observed in these clouds.

To further compare with the thermal Jeans fragmentation model, we show in Fig. \ref{fig:M-separation_hsit} the relationship between the core mass and the separation. For comparison, we plot the core mass and separation normalized to the local Jeans mass ($m_J$) and Jeans length ($l_J$), where $m_J$ and $l_J$ are estimated in the trunks that include the corresponding cores (leaves).
If the cores form by thermal Jeans fragmentation, the cores should be preferentially distributed in the upper right part of $M_{\rm core}/m_J \ge 1$ and $S_{\rm 3D}/l_J \ge 1$.  
There are no such cores found in that part.
Therefore, 
thermal Jeans fragmentation seems to play a minor role in the core formation in these low-mass clouds.

For turbulent Jeans fragmentation, both the characteristic length scale and the mass scale become larger than those of thermal Jeans fragmentation, since the internal cloud turbulence contribute to additional dynamical support.
The turbulent Mach numbers for both clouds are measured to be $\sim 10$  \citep{Spilker_2022,Dame_2001}, resulting in the turbulent Jeans length being about three times greater than the thermal Jeans length.
Therefore, the turbulent Jeans fragmentation is also in disagreement with the observed core properties.

Gravitational fragmentation is strongly influenced by cloud geometry. In the case of filamentary clouds, dense cores typically form with separations around four times the filament width, which is generally several times longer than the thermal Jeans length \citep[see][and references therein]{Ishihara_2024}. Given the widely accepted filament width of 0.1 pc, this would predict core separations of about 0.4 pc—larger than the observed values in our analysis. Therefore, our findings contradict gravitational fragmentation scenarios and suggest that this mechanism is unlikely to explain the core formation in these clouds.

\begin{figure}[htb!]
%\begin{figure*}[H]
\centering
\includegraphics[width=0.82\linewidth]{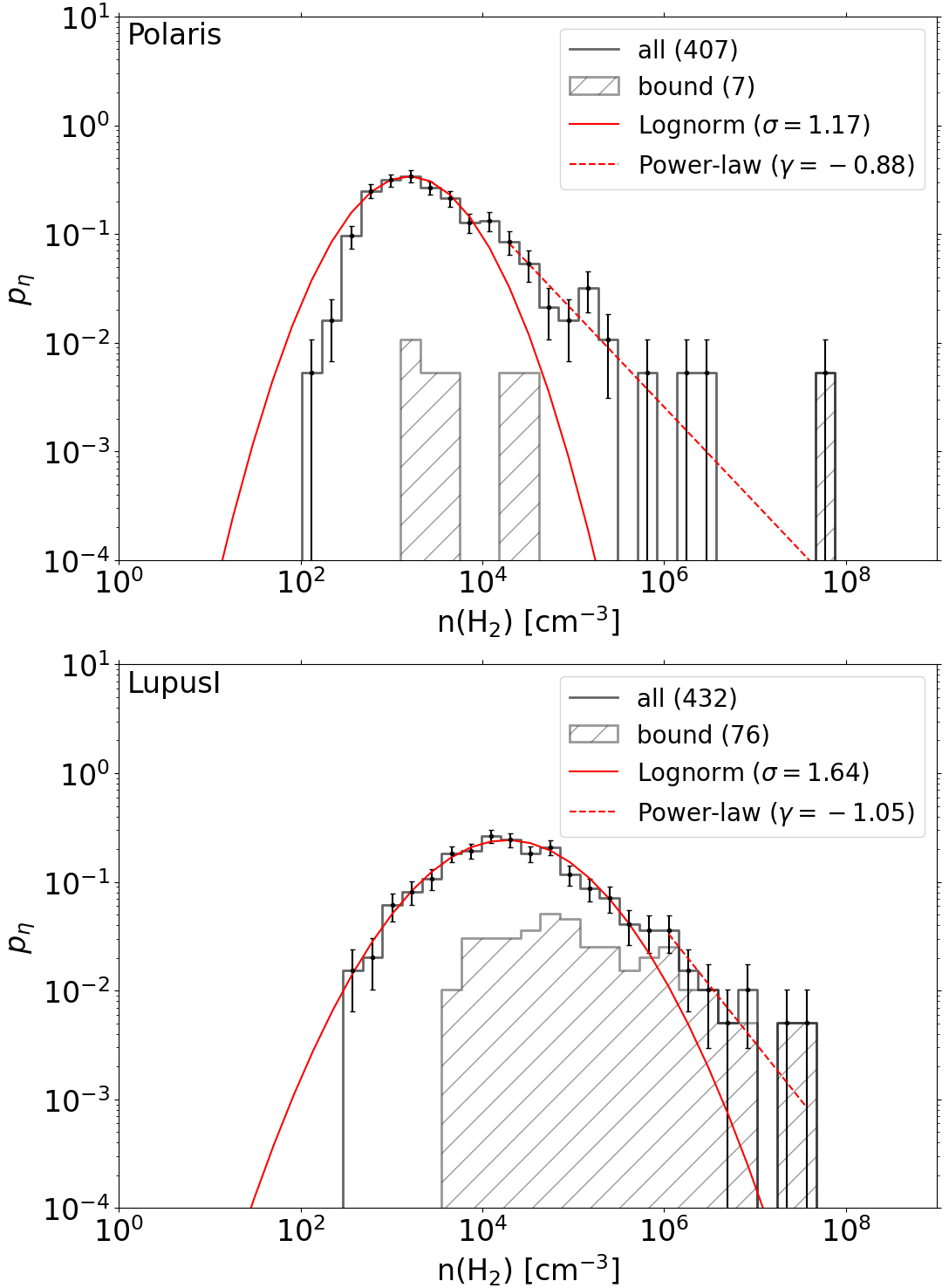}
\caption{Histograms of mean volume density of identified core for Polaris ({\it top}) and Lupus I (${\it bottom}$).
The white and diagonal bins represent the sample of all cores and only bound cores, respectively.
The solid red curves and dotted red line indicate the results of the log-normal fit and power-law fit. The error bar was calculated by Poisson statistics.}
\label{fig:densityPDF}
\end{figure}

\section{Comparison with turbulent fragmentation}
\label{sec:turb}

\subsection{Core separation and mass}

In a turbulent medium, the energy power spectrum can be characterized by a power-law distribution.
In such scale-free distribution, there are no characteristic length scales.
However, in actual cloud turbulence, the thermal width becomes dominant over the nonthermal width at smaller scales. 
The scale below which the nonthermal width coincides with the thermal width is called a sonic scale, $l_S$, \citep{goodman98,Federrath_2012,federrath21}, and 
below $l_S$, the thermal pressure becomes larger than the turbulent pressure.

Thus, the characteristic length scale of turbulent fragmentation corresponds to the cloud's sonic scale, which has been estimated to be about 0.24 pc ($\sim 47000 $ au) based on CO observations for the Polaris \citep{Ossenkopf_2002}. This agrees well with the observed peak core separation of the Polaris.
For Lupus I, there are no precise measurements from current observations, but the sonic scales in nearby clouds with similar environments (e.g. Taurus, Ophiuchus, and Orion) have been measured to be about 0.1 pc 
\citep{Brunt_2010,yun21a,yun21b}.
Therefore, the sonic scale of Lupus I is likely close to 0.1 pc \citep[see also Eq. (13) of ][]{Federrath_2012}.

If the gas is accumulated from the surrounding area within $l_S/2$, 
the characteristic mass scale is given as
\begin{equation}
    M_S = 4\pi \left<\rho\right>(l_S/2) ^3/3  \ 
,\end{equation}
where $\left<\rho\right>$ 
($\approx 3.2 \times 10^2 $ cm$^{-3}$ for Polaris and $\approx 3.0 \times 10^3$ cm$^{-3}$ for Lupus I) is estimated in the main branches identified by \texttt{astrodendro}.
This mass scale is estimated to be 0.16 $M_\odot$ for Polaris and 0.1 M$_\odot$ for Lupus I, which agrees with those of the observations by a factor of a few.
Therefore, we conclude that the core separation and mass are in good agreement with the prediction of turbulent fragmentation.

\subsection{Core density PDFs}

Another important prediction of turbulent fragmentation is its density PDF, which follows a log-normal form. The dispersion of the PDF was determined by the Mach number of cloud turbulence (${\cal M}$) as:
\begin{equation} 
\sigma_{\rho}^2 = \ln (1 + {\cal M}^2 b^2),
\label{eq:dispersion} 
\end{equation}
where $b$ is a constant representing the ratio between compressible and solenoidal components, typically ranging from $b\approx 1/3-1$ \citep{Federrath_2012}.

To investigate the effect of cloud turbulence, we calculated the density PDFs for both clouds.
The density PDFs are sometimes modelled as a combination of a log-normal distribution and a power-law function at higher densities \citep{Spilker_2022, Schneider_2022}, with the power-law tail presumably influenced by cloud self-gravity. 
Here, we applied the fitting method performed by \citet{Schneider_2015_1}.
In \citet{Schneider_2015_1}, the deviation point where the log-normal distribution transitions into a power-law distribution is defined as the residual between the fitted log-normal distribution and the measured PDF being greater than three times the statistical noise in each bin of the PDF.
However, the sample size in this study was small and the statistical noise was large, so the conditions were relaxed from three times to one.

Core densities were calculated under the assumption of spherical symmetry. The resulting density PDFs are plotted in Fig. \ref{fig:densityPDF}. 
The peak densities were estimated to be $1.7\times10^3$ cm$^{-3}$ and $1.9\times10^4$ cm$^{-3}$ for Polaris and Lupus I,  respectively.
The density PDFs of both clouds seem to be well fitted by single log-normal functions, having high-density tails. 
The dispersions of these density PDFs 
%($\le 10^5$ cm$^{-3}$) 
are measured as $\sigma_\rho = $ 1.09 for Polaris, and 1.64 for Lupus I, respectively.
Using Eq. (\ref{eq:dispersion}), we can deduce the expected turbulent Mach number for these clouds.
The expected Mach numbers calculated are ${\cal M} = 3.8-5.1$ and $9.3-12.4$ with $b=1/3-0.4$ 
\citep[solenoidal dominant forcing -- mixing of solenoidal and compressive forcing;][]{Federrath_2012}
%(solenoidal dominant forcing -- mixing of solenoidal and compressive forcing; \citet{Federrath_2012})
for Polaris and Lupus I, respectively.
For Polaris, \citet{federrath10} analyzed the density PDF, suggesting that the solenoidal component is dominant for the Polaris cloud, and this implies that $b\approx 1/3$ may be reasonable.
The turbulent Mach numbers derived from molecular line observations are in reasonable agreement with the deduced Mach numbers. 
\citet{Spilker_2022} derived the Mach number using the CO($J=1-0$) data by \citet{Dame_2001}: ${\cal M} = 12$ for Polaris and ${\cal M} = 11$ for Lupus I.
%\vspace{-2mm}

\section{Conclusion}
\label{sec:conclusion}

To understand the physics behind this process, we identified dense cores and derived their masses and separations in the two nearby low-mass clouds, Polaris and Lupus I.  These clouds are less dense, and thus the effect of self-gravity is less important than in high-mass star-forming clumps.

The most prevalent cloud fragmentation mechanism is gravitational fragmentation, which can be divided into two types: thermal Jeans fragmentation and turbulent Jeans fragmentation. In thermal Jeans fragmentation, the characteristic length and mass scales are the Jeans length ($l_J$) and Jeans mass ($m_J$).
In turbulent Jeans fragmentation, internal nonthermal motion's support is included in the effective sound speed as $
c_s ^{\rm eff} = c_s \sqrt{ 1+ {\cal M}^2} $, 
leading to larger characteristic length and mass scales than those of thermal Jeans fragmentation.

Recent studies have emphasized gravitational fragmentation of filaments as a primary mode of core formation \citep{Andre_2010}. The fragmentation scale or separation for filament fragmentation is about four times the filament width, generally a few times longer than the thermal Jeans length \citep{Ishihara_2024}.
Therefore, the gravitational fragmentation scenarios -- turbulent Jeans fragmentation and filament fragmentation -- predict core masses and separations larger than those of thermal Jeans fragmentation.

Contrary to these predictions, we found that the core masses and separations in both Polaris and Lupus I are smaller than those expected from thermal Jeans fragmentation. 
The peak of core separation appears to be comparable to the sonic scale of $\sim$ 0.1 pc, below which thermal support becomes dominant, in both clouds. 
In addition, the density PDFs constructed from the core's volume densities are fitted well, with log-normal distributions having dispersions estimated from the turbulent Mach numbers.
These observational facts strongly indicate that the cores preferentially form by turbulent fragmentation.

In high-mass star-forming clumps, \citet{Ishihara_2024} found that core separation is comparable to the thermal Jeans length, concluding that core formation is likely controlled by thermal Jeans fragmentation. This apparent difference between high-mass clumps and nearby low-mass clouds may suggest that different physical processes are responsible for core formation in different environments. Alternatively, the higher densities of high-mass star-forming clumps could mean that the characteristic mass scale at the sonic scale, $M_S$, may already exceed or be at least comparable to the thermal Jeans mass. 
If this were the case, turbulent fragmentation might still be the primary driver of core formation, even in high-mass clumps.
The sonic scales in some high-mass star-forming clumps are measured to be $\sim$ 0.01 pc ($\approx 2\times 10^3$ au), about ten times smaller than the ones of nearby molecular clouds \citep{ohashi16,li23}.
These smaller sonic scales appear to be consistent with the core separation measured in high-mass star-forming clumps within a factor of a few \citep{Ishihara_2024}.
More precise measurements of the sonic scale in these regions could provide further insights into understanding the cloud fragmentation process in various environments.

\begin{acknowledgements}
This work was supported in part by The Graduate University for Advanced Studies, SOKENDAI. This work was financially supported by JSPS KAKENHI Grant Numbers JP22H01271 (PS), JP23H01218 (FN), and JP23H01221 (PS).
\end{acknowledgements}
%\vspace{-4mm}

\bibliographystyle{aa} % style aa.bst
\bibliography{references,reference2}

\begin{thebibliography}{43}
\expandafter\ifx\csname natexlab\endcsname\relax\def\natexlab#1{#1}\fi

\bibitem[{{Andr{\'e}} {et~al.}(2014){Andr{\'e}}, {Di Francesco},
  {Ward-Thompson}, {Inutsuka}, {Pudritz}, \& {Pineda}}]{Andre_2014}
{Andr{\'e}}, P., {Di Francesco}, J., {Ward-Thompson}, D., {et~al.} 2014, in
  Protostars and Planets VI, ed. H.~{Beuther}, R.~S. {Klessen}, C.~P.
  {Dullemond}, \& T.~{Henning}, 27

\bibitem[{{Andr{\'e}} {et~al.}(2010){Andr{\'e}}, {Men'shchikov}, {Bontemps},
  {K{\"o}nyves}, {Motte}, {Schneider}, {Didelon}, {Minier}, {Saraceno},
  {Ward-Thompson}, {di Francesco}, {White}, {Molinari}, {Testi}, {Abergel},
  {Griffin}, {Henning}, {Royer}, {Mer{\'\i}n}, {Vavrek}, {Attard},
  {Arzoumanian}, {Wilson}, {Ade}, {Aussel}, {Baluteau}, {Benedettini},
  {Bernard}, {Blommaert}, {Cambr{\'e}sy}, {Cox}, {di Giorgio}, {Hargrave},
  {Hennemann}, {Huang}, {Kirk}, {Krause}, {Launhardt}, {Leeks}, {Le Pennec},
  {Li}, {Martin}, {Maury}, {Olofsson}, {Omont}, {Peretto}, {Pezzuto}, {Prusti},
  {Roussel}, {Russeil}, {Sauvage}, {Sibthorpe}, {Sicilia-Aguilar}, {Spinoglio},
  {Waelkens}, {Woodcraft}, \& {Zavagno}}]{Andre_2010}
{Andr{\'e}}, P., {Men'shchikov}, A., {Bontemps}, S., {et~al.} 2010, \aap, 518,
  L102

\bibitem[{{Benedettini} {et~al.}(2018){Benedettini}, {Pezzuto}, {Schisano},
  {Andr{\'e}}, {K{\"o}nyves}, {Men'shchikov}, {Ladjelate}, {Di Francesco},
  {Elia}, {Arzoumanian}, {Louvet}, {Palmeirim}, {Rygl}, {Schneider},
  {Spinoglio}, \& {Ward-Thompson}}]{Benedettini_2018}
{Benedettini}, M., {Pezzuto}, S., {Schisano}, E., {et~al.} 2018, \aap, 619, A52

\bibitem[{{Beuther} {et~al.}(2022){Beuther}, {Schneider}, {Simon}, {Suri},
  {Ossenkopf-Okada}, {Kabanovic}, {R{\"o}llig}, {Guevara}, {Tielens},
  {Sandell}, {Buchbender}, {Ricken}, \& {G{\"u}sten}}]{Beuther_2022}
{Beuther}, H., {Schneider}, N., {Simon}, R., {et~al.} 2022, \aap, 659, A77

\bibitem[{{Brunt}(2010)}]{Brunt_2010}
{Brunt}, C.~M. 2010, \aap, 513, A67

\bibitem[{{Dame} {et~al.}(2001){Dame}, {Hartmann}, \& {Thaddeus}}]{Dame_2001}
{Dame}, T.~M., {Hartmann}, D., \& {Thaddeus}, P. 2001, \apj, 547, 792

\bibitem[{{Federrath} \& {Klessen}(2012)}]{Federrath_2012}
{Federrath}, C. \& {Klessen}, R.~S. 2012, \apj, 761, 156

\bibitem[{{Federrath} {et~al.}(2021){Federrath}, {Klessen}, {Iapichino}, \&
  {Beattie}}]{federrath21}
{Federrath}, C., {Klessen}, R.~S., {Iapichino}, L., \& {Beattie}, J.~R. 2021,
  Nature Astronomy, 5, 365

\bibitem[{{Federrath} {et~al.}(2010){Federrath}, {Roman-Duval}, {Klessen},
  {Schmidt}, \& {Mac Low}}]{federrath10}
{Federrath}, C., {Roman-Duval}, J., {Klessen}, R.~S., {Schmidt}, W., \& {Mac
  Low}, M.~M. 2010, \aap, 512, A81

\bibitem[{{Goodman} {et~al.}(1998){Goodman}, {Barranco}, {Wilner}, \&
  {Heyer}}]{goodman98}
{Goodman}, A.~A., {Barranco}, J.~A., {Wilner}, D.~J., \& {Heyer}, M.~H. 1998,
  \apj, 504, 223

\bibitem[{{Ishihara} {et~al.}(2024){Ishihara}, {Sanhueza}, {Nakamura}, {Saito},
  {Chen}, {Li}, {Olguin}, {Taniguchi}, {Morii}, {Lu}, {Luo}, {Sakai}, \&
  {Zhang}}]{Ishihara_2024}
{Ishihara}, K., {Sanhueza}, P., {Nakamura}, F., {et~al.} 2024, \apj, 974, 95

\bibitem[{{Jeans}(1928)}]{Jeans_1928}
{Jeans}, J.~H. 1928, {Astronomy and cosmogony} (Cambridge University Press)

\bibitem[{{Kirk} {et~al.}(2017){Kirk}, {Friesen}, {Pineda}, {Rosolowsky},
  {Offner}, {Matzner}, {Myers}, {Di Francesco}, {Caselli}, {Alves},
  {Chac{\'o}n-Tanarro}, {Chen}, {Chun-Yuan Chen}, {Keown}, {Punanova}, {Seo},
  {Shirley}, {Ginsburg}, {Hall}, {Singh}, {Arce}, {Goodman}, {Martin}, \&
  {Redaelli}}]{Kirk_2017}
{Kirk}, H., {Friesen}, R.~K., {Pineda}, J.~E., {et~al.} 2017, \apj, 846, 144

\bibitem[{{Klaassen} {et~al.}(2018){Klaassen}, {Johnston}, {Urquhart},
  {Mottram}, {Peters}, {Kuiper}, {Beuther}, {van der Tak}, \&
  {Goddi}}]{klaassen18}
{Klaassen}, P.~D., {Johnston}, K.~G., {Urquhart}, J.~S., {et~al.} 2018, \aap,
  611, A99

\bibitem[{{Knuth}(1973)}]{Knuth_1973}
{Knuth}, D.~E. 1973, {The art of computer programming. Vol.3: Sorting and
  searching} (Addison-Wesley Professional)

\bibitem[{{K{\"o}nyves} {et~al.}(2010){K{\"o}nyves}, {Andr{\'e}},
  {Men'shchikov}, {Schneider}, {Arzoumanian}, {Bontemps}, {Attard}, {Motte},
  {Didelon}, {Maury}, {Abergel}, {Ali}, {Baluteau}, {Bernard}, {Cambr{\'e}sy},
  {Cox}, {di Francesco}, {di Giorgio}, {Griffin}, {Hargrave}, {Huang}, {Kirk},
  {Li}, {Martin}, {Minier}, {Molinari}, {Olofsson}, {Pezzuto}, {Russeil},
  {Roussel}, {Saraceno}, {Sauvage}, {Sibthorpe}, {Spinoglio}, {Testi},
  {Ward-Thompson}, {White}, {Wilson}, {Woodcraft}, \& {Zavagno}}]{Konyves_2010}
{K{\"o}nyves}, V., {Andr{\'e}}, P., {Men'shchikov}, A., {et~al.} 2010, \aap,
  518, L106

\bibitem[{{Larson}(1985)}]{Larson_1985}
{Larson}, R.~B. 1985, \mnras, 214, 379

\bibitem[{{Li} {et~al.}(2023){Li}, {Sanhueza}, {Zhang}, {Guido}, {Sabatini},
  {Morii}, {Lu}, {Tafoya}, {Nakamura}, {Izumi}, {Tatematsu}, \& {Li}}]{li23}
{Li}, S., {Sanhueza}, P., {Zhang}, Q., {et~al.} 2023, \apj, 949, 109

\bibitem[{{Lu} {et~al.}(2020){Lu}, {Cheng}, {Ginsburg}, {Longmore},
  {Kruijssen}, {Battersby}, {Zhang}, \& {Walker}}]{Lu_2020}
{Lu}, X., {Cheng}, Y., {Ginsburg}, A., {et~al.} 2020, \apjl, 894, L14

\bibitem[{{Mac Low} \& {Klessen}(2004)}]{maclow04}
{Mac Low}, M.-M. \& {Klessen}, R.~S. 2004, Reviews of Modern Physics, 76, 125

\bibitem[{{Maruta} {et~al.}(2010){Maruta}, {Nakamura}, {Nishi}, {Ikeda}, \&
  {Kitamura}}]{maruta10}
{Maruta}, H., {Nakamura}, F., {Nishi}, R., {Ikeda}, N., \& {Kitamura}, Y. 2010,
  \apj, 714, 680

\bibitem[{{Men'shchikov} {et~al.}(2010){Men'shchikov}, {Andr{\'e}}, {Didelon},
  {K{\"o}nyves}, {Schneider}, {Motte}, {Bontemps}, {Arzoumanian}, {Attard},
  {Abergel}, {Baluteau}, {Bernard}, {Cambr{\'e}sy}, {Cox}, {di Francesco}, {di
  Giorgio}, {Griffin}, {Hargrave}, {Huang}, {Kirk}, {Li}, {Martin}, {Minier},
  {Miville-Desch{\^e}nes}, {Molinari}, {Olofsson}, {Pezzuto}, {Roussel},
  {Russeil}, {Saraceno}, {Sauvage}, {Sibthorpe}, {Spinoglio}, {Testi},
  {Ward-Thompson}, {White}, {Wilson}, {Woodcraft}, \&
  {Zavagno}}]{menshchikov10}
{Men'shchikov}, A., {Andr{\'e}}, P., {Didelon}, P., {et~al.} 2010, \aap, 518,
  L103

\bibitem[{{Miville-Desch{\^e}nes} {et~al.}(2010){Miville-Desch{\^e}nes},
  {Martin}, {Abergel}, {Bernard}, {Boulanger}, {Lagache}, {Anderson},
  {Andr{\'e}}, {Arab}, {Baluteau}, {Blagrave}, {Bontemps}, {Cohen},
  {Compiegne}, {Cox}, {Dartois}, {Davis}, {Emery}, {Fulton}, {Gry}, {Habart},
  {Huang}, {Joblin}, {Jones}, {Kirk}, {Lim}, {Madden}, {Makiwa}, {Menshchikov},
  {Molinari}, {Moseley}, {Motte}, {Naylor}, {Okumura}, {Pinheiro
  Gon{\c{c}}alves}, {Polehampton}, {Rod{\'o}n}, {Russeil}, {Saraceno},
  {Schneider}, {Sidher}, {Spencer}, {Swinyard}, {Ward-Thompson}, {White}, \&
  {Zavagno}}]{miville10}
{Miville-Desch{\^e}nes}, M.~A., {Martin}, P.~G., {Abergel}, A., {et~al.} 2010,
  \aap, 518, L104

\bibitem[{{Morii} {et~al.}(2024){Morii}, {Sanhueza}, {Zhang}, {Nakamura}, {Li},
  {Sabatini}, {Olguin}, {Beuther}, {Tafoya}, {Izumi}, {Tatematsu}, \&
  {Sakai}}]{morii24}
{Morii}, K., {Sanhueza}, P., {Zhang}, Q., {et~al.} 2024, \apj, 966, 171

\bibitem[{{Ohashi} {et~al.}(2016){Ohashi}, {Sanhueza}, {Chen}, {Zhang},
  {Busquet}, {Nakamura}, {Palau}, \& {Tatematsu}}]{ohashi16}
{Ohashi}, S., {Sanhueza}, P., {Chen}, H.-R.~V., {et~al.} 2016, \apj, 833, 209

\bibitem[{{Ossenkopf} \& {Mac Low}(2002)}]{Ossenkopf_2002}
{Ossenkopf}, V. \& {Mac Low}, M.~M. 2002, \aap, 390, 307

\bibitem[{{Padoan} \& {Nordlund}(2002)}]{padoan02}
{Padoan}, P. \& {Nordlund}, {\r{A}}. 2002, \apj, 576, 870

\bibitem[{{Padoan} {et~al.}(2020){Padoan}, {Pan}, {Juvela}, {Haugb{\o}lle}, \&
  {Nordlund}}]{Padoan_2020}
{Padoan}, P., {Pan}, L., {Juvela}, M., {Haugb{\o}lle}, T., \& {Nordlund},
  {\r{A}}. 2020, \apj, 900, 82

\bibitem[{{Palau} {et~al.}(2015){Palau}, {Ballesteros-Paredes},
  {V{\'a}zquez-Semadeni}, {S{\'a}nchez-Monge}, {Estalella}, {Fall}, {Zapata},
  {Camacho}, {G{\'o}mez}, {Naranjo-Romero}, {Busquet}, \& {Fontani}}]{palau15}
{Palau}, A., {Ballesteros-Paredes}, J., {V{\'a}zquez-Semadeni}, E., {et~al.}
  2015, \mnras, 453, 3785

\bibitem[{{Panopoulou} {et~al.}(2022){Panopoulou}, {Clark}, {Hacar}, {Heitsch},
  {Kainulainen}, {Ntormousi}, {Seifried}, \& {Smith}}]{panopoulou22}
{Panopoulou}, G.~V., {Clark}, S.~E., {Hacar}, A., {et~al.} 2022, \aap, 657, L13

\bibitem[{{Robitaille} {et~al.}(2019){Robitaille}, {Motte}, {Schneider},
  {Elia}, \& {Bontemps}}]{robitaille19}
{Robitaille}, J.~F., {Motte}, F., {Schneider}, N., {Elia}, D., \& {Bontemps},
  S. 2019, \aap, 628, A33

\bibitem[{{Rosolowsky} {et~al.}(2008){Rosolowsky}, {Pineda}, {Kauffmann}, \&
  {Goodman}}]{Rosolowsky_2008}
{Rosolowsky}, E.~W., {Pineda}, J.~E., {Kauffmann}, J., \& {Goodman}, A.~A.
  2008, \apj, 679, 1338

\bibitem[{{Sanhueza} {et~al.}(2019){Sanhueza}, {Contreras}, {Wu}, {Jackson},
  {Guzm{\'a}n}, {Zhang}, {Li}, {Lu}, {Silva}, {Izumi}, {Liu}, {Miura},
  {Tatematsu}, {Sakai}, {Beuther}, {Garay}, {Ohashi}, {Saito}, {Nakamura},
  {Saigo}, {Veena}, {Nguyen-Luong}, \& {Tafoya}}]{sanhueza19}
{Sanhueza}, P., {Contreras}, Y., {Wu}, B., {et~al.} 2019, \apj, 886, 102

\bibitem[{{Schneider} {et~al.}(2015){Schneider}, {Ossenkopf}, {Csengeri},
  {Klessen}, {Federrath}, {Tremblin}, {Girichidis}, {Bontemps}, \&
  {Andr{\'e}}}]{Schneider_2015_1}
{Schneider}, N., {Ossenkopf}, V., {Csengeri}, T., {et~al.} 2015, \aap, 575, A79

\bibitem[{{Schneider} {et~al.}(2022){Schneider}, {Ossenkopf-Okada}, {Clarke},
  {Klessen}, {Kabanovic}, {Veltchev}, {Bontemps}, {Dib}, {Csengeri},
  {Federrath}, {Di Francesco}, {Motte}, {Andr{\'e}}, {Arzoumanian}, {Beattie},
  {Bonne}, {Didelon}, {Elia}, {K{\"o}nyves}, {Kritsuk}, {Ladjelate}, {Myers},
  {Pezzuto}, {Robitaille}, {Roy}, {Seifried}, {Simon}, {Soler}, \&
  {Ward-Thompson}}]{Schneider_2022}
{Schneider}, N., {Ossenkopf-Okada}, V., {Clarke}, S., {et~al.} 2022, \aap, 666,
  A165

\bibitem[{{Shimajiri} {et~al.}(2015){Shimajiri}, {Kitamura}, {Nakamura},
  {Momose}, {Saito}, {Tsukagoshi}, {Hiramatsu}, {Shimoikura}, {Dobashi},
  {Hara}, \& {Kawabe}}]{shimajiri15}
{Shimajiri}, Y., {Kitamura}, Y., {Nakamura}, F., {et~al.} 2015, \apjs, 217, 7

\bibitem[{{Spilker} {et~al.}(2022){Spilker}, {Kainulainen}, \&
  {Orkisz}}]{Spilker_2022}
{Spilker}, A., {Kainulainen}, J., \& {Orkisz}, J. 2022, \aap, 667, A110

\bibitem[{{Takemura} {et~al.}(2023){Takemura}, {Nakamura}, {Arce}, {Schneider},
  {Ossenkopf-Okada}, {Kong}, {Ishii}, {Dobashi}, {Shimoikura}, {Sanhueza},
  {Tsukagoshi}, {Padoan}, {Klessen}, {Goldsmith}, {Burkhart}, {Lis},
  {S{\'a}nchez-Monge}, {Shimajiri}, \& {Kawabe}}]{takemura23}
{Takemura}, H., {Nakamura}, F., {Arce}, H.~G., {et~al.} 2023, \apjs, 264, 35

\bibitem[{{Takemura} {et~al.}(2021){Takemura}, {Nakamura}, {Ishii},
  {Shimajiri}, {Sanhueza}, {Tsukagoshi}, {Kawabe}, {Hirota}, \&
  {Kataoka}}]{takemura21b}
{Takemura}, H., {Nakamura}, F., {Ishii}, S., {et~al.} 2021, \pasj, 73, 487
  (Paper I)

\bibitem[{{Ward-Thompson} {et~al.}(2010){Ward-Thompson}, {Kirk}, {Andr{\'e}},
  {Saraceno}, {Didelon}, {K{\"o}nyves}, {Schneider}, {Abergel}, {Baluteau},
  {Bernard}, {Bontemps}, {Cambr{\'e}sy}, {Cox}, {di Francesco}, {di Giorgio},
  {Griffin}, {Hargrave}, {Huang}, {Li}, {Martin}, {Men'shchikov}, {Minier},
  {Molinari}, {Motte}, {Olofsson}, {Pezzuto}, {Russeil}, {Sauvage},
  {Sibthorpe}, {Spinoglio}, {Testi}, {White}, {Wilson}, {Woodcraft}, \&
  {Zavagno}}]{ward-thompson10}
{Ward-Thompson}, D., {Kirk}, J.~M., {Andr{\'e}}, P., {et~al.} 2010, \aap, 518,
  L92

\bibitem[{{Yun} {et~al.}(2021{\natexlab{a}}){Yun}, {Lee}, {Choi}, {Evans},
  {Offner}, {Heyer}, {Gaches}, {Lee}, {Baek}, {Choi}, {Kang}, {Lee},
  {Tatematsu}, {Yang}, {Chen}, {Lee}, {Jung}, {Lee}, \& {Cho}}]{yun21b}
{Yun}, H.-S., {Lee}, J.-E., {Choi}, Y., {et~al.} 2021{\natexlab{a}}, \apjs,
  256, 16

\bibitem[{{Yun} {et~al.}(2021{\natexlab{b}}){Yun}, {Lee}, {Evans}, {Offner},
  {Heyer}, {Cho}, {Gaches}, {Yang}, {Chen}, {Choi}, {Lee}, {Baek}, {Choi},
  {Kim}, {Kang}, {Lee}, \& {Tatematsu}}]{yun21a}
{Yun}, H.-S., {Lee}, J.-E., {Evans}, N.~J., {et~al.} 2021{\natexlab{b}}, \apj,
  921, 31

\bibitem[{{Zhang} {et~al.}(2022){Zhang}, {Evans}, {Liu}, {Wu}, {Wang}, {Liu},
  {Zhu}, {Ren}, {Dewangan}, {Lee}, {Li}, {Bronfman}, {Tej}, \&
  {Li}}]{Zhang_C_2022_ATOMS}
{Zhang}, C., {Evans}, N.~J., {Liu}, T., {et~al.} 2022, \mnras, 510, 4998

\end{thebibliography}
% your 

%\clearpage

%%%%%%%%%%%%%%%%%%%%%%%%%%%%%%%%%%%%%%%%%%%%%%%%%%%%%%%%%%%%%%%%%%%%%%%%%%%%%%%%

\begin{appendix} %First appendix

\section{Core distribution in the clouds}
\label{sec:clouds}

\begin{figure}[htb!]
\centering
\includegraphics[angle=0,width=\linewidth]{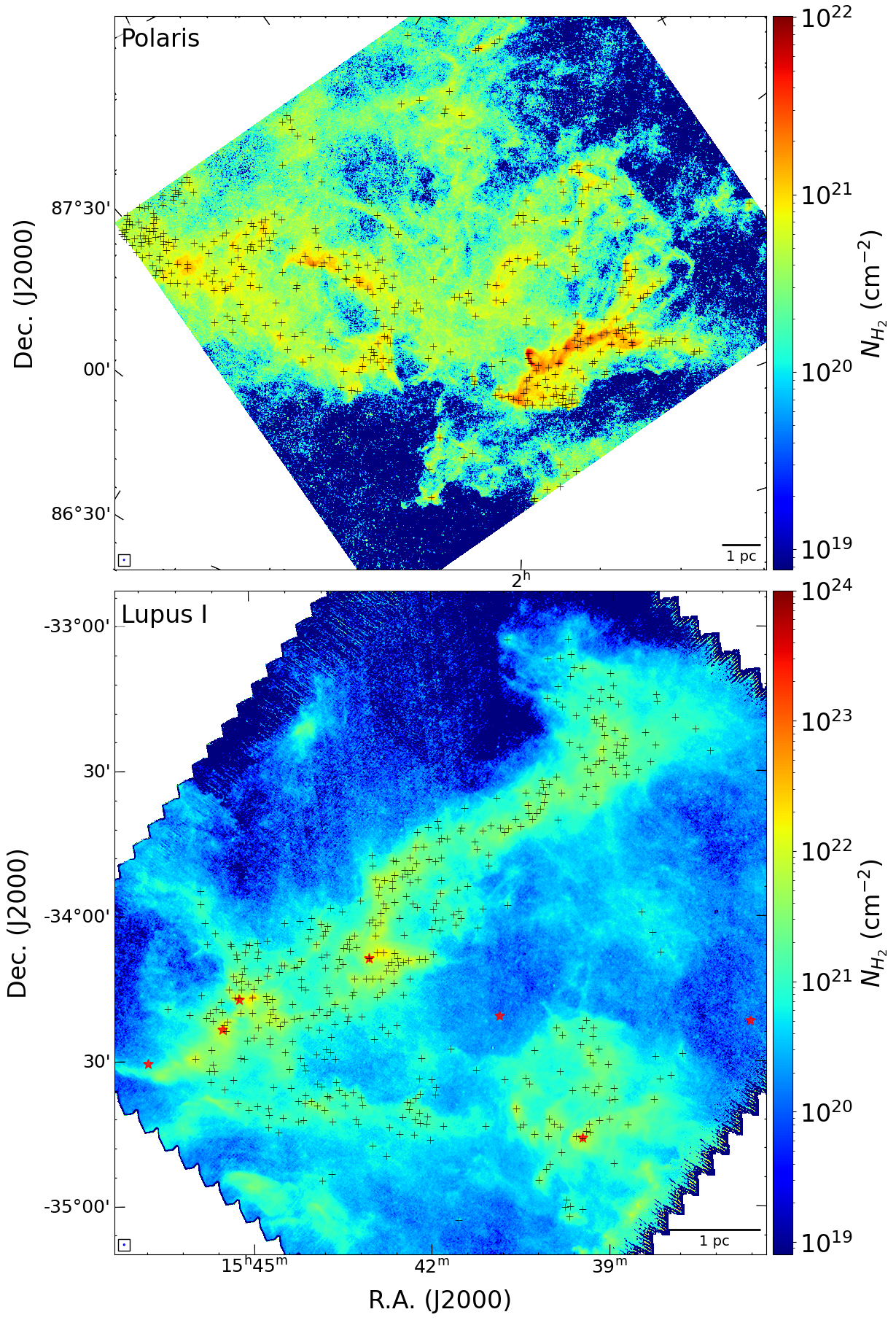}
  \caption{Column density distributions of target clouds. The H$_2$ column density maps of Polaris ({\it top}) and Lupus I ({\it bottom}).
  Red star and black cross symbols are the positions of the protostellar cores identified by \citep{Benedettini_2018} and those of the dense cores identified by \texttt{astrodendro} (this work).
  }
\label{fig:herschel maps}
\end{figure}

Figs. \ref{fig:herschel maps}a and \ref{fig:herschel maps}b show the column density maps of Polaris and Lupus I, respectively. For the Polaris, we trimmed the original map to account for variations in background levels and noise across different regions. In the trimmed map, we used \texttt{astrodendro} to identify the cores. The identified cores are marked with crosses in each panel.
For Lupus I, we excluded $\sim$ 24\arcmin \, wide parts from the edges of the map due to the prominence of (artificial) scan patterns.

\section{Distances to the targets}
\label{sec:distance}

The target distances are recently updated by the Gaia data. In this paper, we adopted the distances, 355 pc and 182 pc, calculated by \citet{panopoulou22} for Polaris and \citet{Benedettini_2018} for Lupus I, respectively. These distances are based on the Gaia data. 
For the Polaris distance, one of the positions in the Appendix of \citet{panopoulou22} ($l,b$)=(123.7, 24.8), overlaps with the Herschel map at 
(RA, Dec)=(01h58m11s,87d34m17s).  Therefore, we adopted 355 pc.

\section{Calculation of map noise levels}
\label{sec:comparison}

In our analysis, we calculated the background emission to be $3.80 \times 10^{20}$ cm$^{-2}$ $\pm$ 7.67 $\times 10^{19}$ cm$^{-2}$ for Polaris and $2.16 \times 10^{20}$ cm$^{-2}$ $\pm$ 8.02 $\times 10^{19}$ cm$^{-2}$ for Lupus I, respectively, based on the areas shown in Fig. \ref{fig:herschel maps}.
For core identification, we subtracted these background components from the column density maps. The \texttt{astrodendro} algorithm was then applied to the processed maps, using the calculated standard deviations as the 1-$\sigma$ noise levels. When estimating core masses, we added the subtracted background component back to the core mass. Notably, this background subtraction does not significantly influence the analysis presented in this paper.

The Herschel maps are also subject to cosmic infrared background (CIB) emission from unresolved high-redshift infrared galaxies \citep{miville10, robitaille19}. As a result, some of the identified cores could correspond to such sources. However, we consider the fraction of these "fake" cores to be negligible in our analysis.

It is worth mentioning that \citet{ward-thompson10} identified only five cores in the Polaris. Their results agree with ours if we adopt their 1-$\sigma$ threshold of $\sim 1.2 \times 10^{21}$ cm$^{-2}$. However, \citet{Andre_2010} estimated the 1$\sigma$ noise level of the Polaris image to be much lower, around $6 \times 10^{19}$ cm$^{-2}$, due to the significantly low background contamination in the Polaris region. This discrepancy in the number of identified cores is likely due to the different 1$\sigma$ noise levels used in these studies, with many more cores being detected at the lower threshold.
In fact, \citet{menshchikov10} reported $\sim 300$ starless cores in Polaris.
If we adopt 1$\sigma \sim 1\times 10^{21}$  cm$^{-2}$, our identified core number becomes $\sim$ 5, the same as that of \citet{ward-thompson10}.
Many our cores in Polaris are distributed along the elongated structures seen in yellow--red in the right panel of Fig. \ref{fig:herschel maps}.

\section{Comparison with the \texttt{getsources} core catalogue}
 \label{sec:dendro}

\begin{figure*}[htb!]
%\begin{figure*}[H]
\centering
\includegraphics[width= \linewidth]{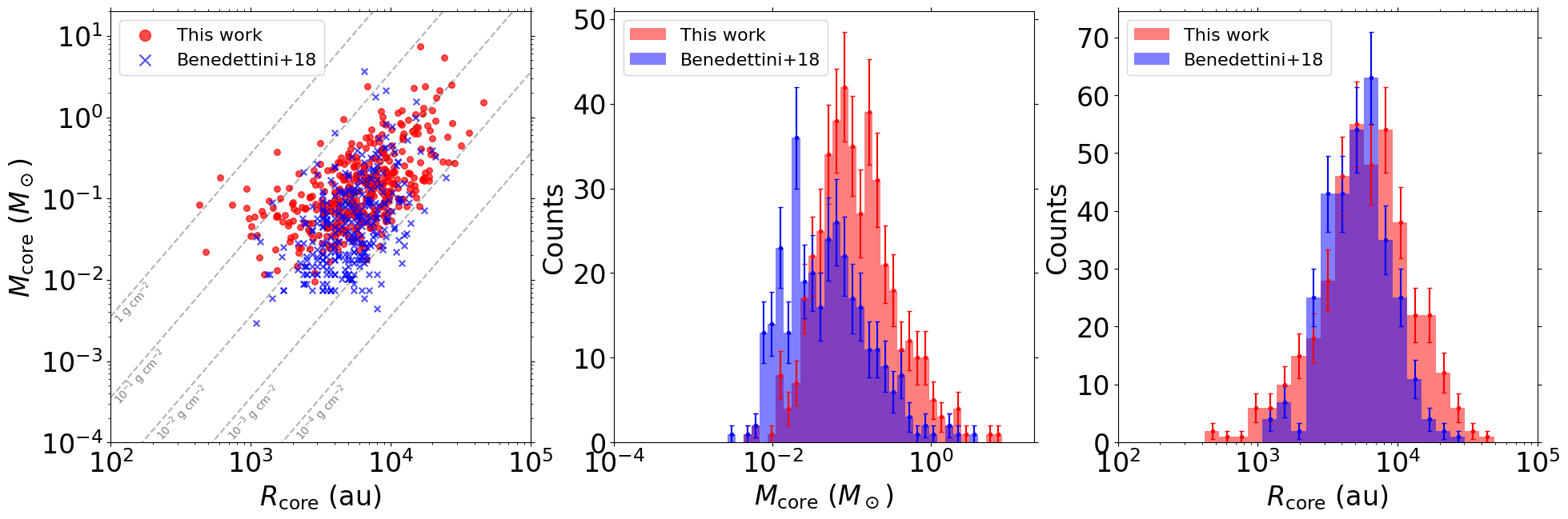}
\caption{Comparison of between \texttt{astrodendro} cores (this paper) and \texttt{getsources} cores \citep{Benedettini_2018}. Core mass versus radius diagram ({\it left}) and histograms of core mass (${\it middle}$) and core radius (${\it right}$) identified by \citet{Benedettini_2018} using \texttt{getsources} and this work using \texttt{astrodendro}. The red circle and blue cross symbols represent the cores by this work and \citet{Benedettini_2018}, respectively.
The error bar was calculated by Poisson statistics.
\label{fig:M-R_hsit_comp}}
\end{figure*}

\begin{figure*}[htb!]
%\begin{figure*}[H]
\centering
\includegraphics[width=0.8 \linewidth]{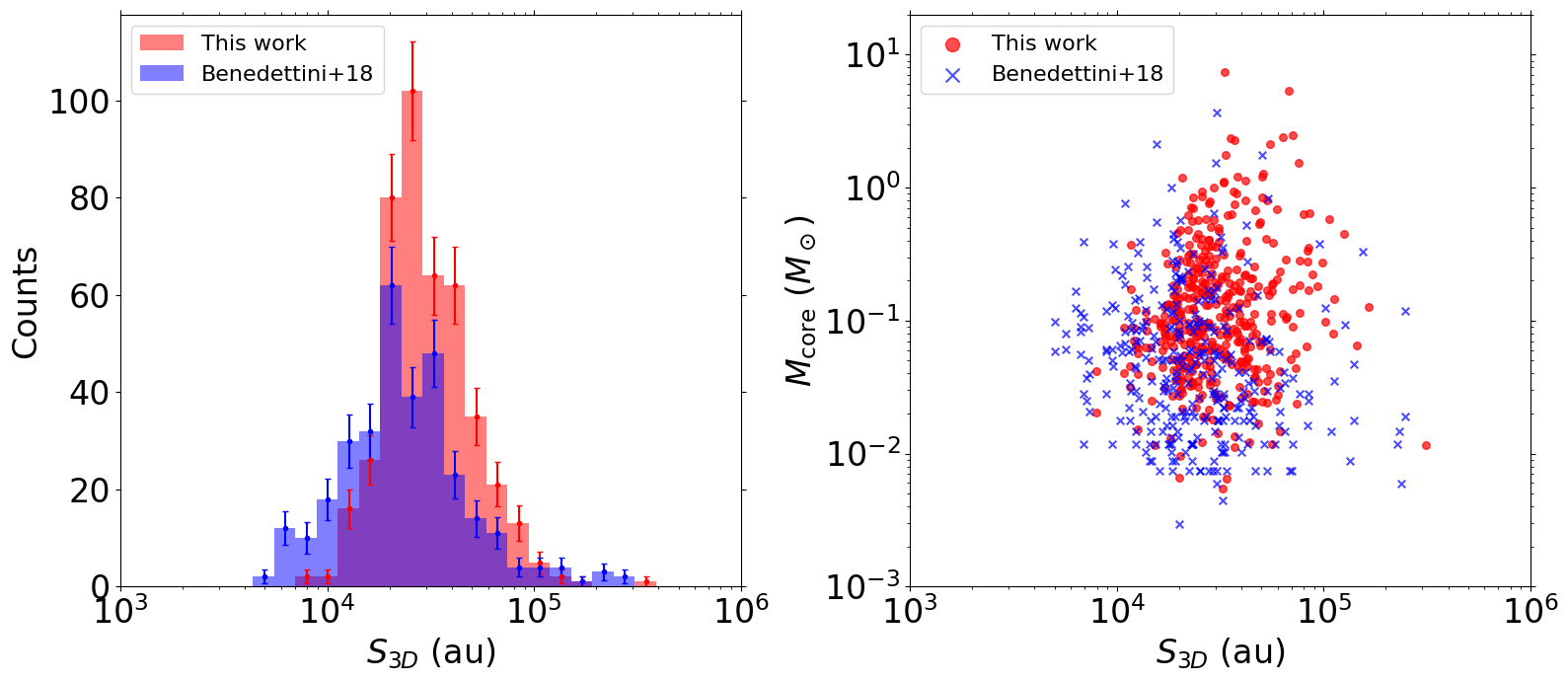}
\caption{Comparison of between \texttt{astrodendro} cores (this paper) and \texttt{getsources} cores \citep{Benedettini_2018}. Histogram of core separation (${\it left}$) and core mass versus separation diagrams in physical scale (${\it right}$).
The red circle and blue cross symbols represent the cores by this work and \citet{Benedettini_2018}, respectively.
The error bar was calculated by Poisson statistics.
\label{fig:M-separation_hsit_comp}}
\end{figure*}

In Figs. \ref{fig:M-R_hsit_comp} and \ref{fig:M-separation_hsit_comp}, we present the core masses, radii, and separations for the Lupus I cloud, comparing results derived using \texttt{astrodendro} (this paper) with those from the core catalogue of \citet{Benedettini_2018}, obtained via \texttt{getsources}.
The mean values and overall distributions show good agreement, indicating that our conclusion regarding the importance of turbulent fragmentation is robust and independent of the core identification method.
Furthermore, the typical cores identified by \texttt{getsources} exhibit smaller values of $\alpha_{\rm BE}$ ($\left<\alpha_{\rm BE}\right>\approx$ 0.25), which is consistent with the dominance of unbound cores (about 80\%), similar to findings in Ophiuchus and Orion A.

\section{Dependence of \texttt{astrodendro} parameter}
\label{sec:dependence}

The \texttt{astrodendro} algorithm relies on three key parameters:
min\_value, min\_delta, and min\_npix.  The third parameter, min\_npix, is typically set to match the angular resolution of the map. To examine the influence of the first two parameters on core properties, we
varied min\_value and min\_delta in the ranges of 3$\sigma$ to 7$\sigma$, and 2$\sigma$ to 4$\sigma$, respectively. The results are briefly summaries in Tab. \ref{tab:comparison}.
Within these parameter ranges, we found that the total number of cores is sensitive to min\_delta, while not to min\_value.
The mean core mass and core separation are also sensitive to min\_delta, but there are no significant differences.
Therefore, we adopted min\_value = 5$\sigma$ and min\_delta = 3$\sigma$ as the representative values in this paper.

\begin{table}%[htb]
\centering
\caption{Physical properties for various dendrogram parameters.\label{tab:dendrogram}}
\begin{tabular}{lcccc}
\hline
Parameter & Cloud & $\mathcal{N}_{\rm core}$ & $\left< M_{\rm core} \right>$ & $\left< S_{\rm 3D} \right>$ \\
(min\_value, min\_delta) &  &  & ($M_\odot$) & (pc) \\
\hline
$(3\sigma, 3\sigma)$ & Polaris & 410 & 0.33 & 0.27 \\
$(5\sigma, 3\sigma)$ & Polaris & 407 & 0.28 & 0.27 \\
$(7\sigma, 3\sigma)$ & Polaris & 388 & 0.21 & 0.27 \\
$(5\sigma, 2\sigma)$ & Polaris & 656 & 0.15 & 0.19 \\
$(5\sigma, 4\sigma)$ & Polaris & 255 & 0.53 & 0.38 \\
$(3\sigma, 3\sigma)$ & Lupus I & 435 & 0.19 & 0.13 \\
$(5\sigma, 3\sigma)$ & Lupus I & 432 & 0.24 & 0.13 \\
$(7\sigma, 3\sigma)$ & Lupus I & 432 & 0.17 & 0.13 \\
$(5\sigma, 2\sigma)$ & Lupus I & 777 & 0.10 & 0.10 \\
$(5\sigma, 4\sigma)$ & Lupus I & 293 & 0.28 & 0.16 \\
\hline
\end{tabular}
\label{tab:comparison}
\end{table}

\section{Nearest neighbour search and the core separation}
\label{sec:nns}

A NNS is a method of finding the closest point to a given query point from a set of data points. As the input data, we used the core center locations in the 2D plane of sky. Using the tool in the matlab library, we computed the distance from the query core center to the nearest neighbour core center. We used the distance as a core separation of the query core.
\citet{Ishihara_2024} derived the correction factor to convert the 2D core separation to 3D value.
Assuming that the observed cores are uniformly distributed in a spherical space,
the 3D core separation is computed as
\begin{equation}
  S_{\rm 3D}  = \frac{S_{\rm obs}}{(4\pi)^{-1} \int ^{\pi/2}_{-\pi/2}\sin \theta \times 2\pi \sin \theta d \theta} = S_{\rm 2D} \times \frac{4}{\pi}
\end{equation}
See \citet{Ishihara_2024} for more details.

\end{appendix}

%references Yourfile.bib

\end{document}